\documentclass{PoS}

\usepackage{graphicx}
\usepackage[square,sort&compress]{natbib}
\usepackage{overpic}
\usepackage{amsmath}
\usepackage{wasysym}
\usepackage{url}
\usepackage[amssymb]{SIunits}
\usepackage{booktabs}
\usepackage{afterpage}

\addunit{\GeV}{\giga\electronvolt}
\addunit{\TeV}{\tera\electronvolt}
\addunit{\eV}{\electronvolt}
\addunit{\smm}{\square\metre\usk\second}
\addunit{\smmr}{\rpsquare\metre\usk\reciprocal\second}

\title{A template method for measuring the iron spectrum in cosmic rays with Cherenkov telescopes }

\ShortTitle{A template method for measuring the iron spectrum with Cherenkov telescopes}

\author{\speaker{Henrike Fleischhack} ~for the VERITAS collaboration\thanks{\texttt{http://veritas.sao.arizona.edu}}\\
        DESY, Platanenallee 6, 15738 Zeuthen, Germany\\
        E-mail: \email{henrike.fleischhack@desy.de}}

\abstract{
The energy-dependent abundance of elements in cosmic rays plays an important role in understanding their acceleration and propagation. Most current results are obtained either from direct measurements by balloon- or satellite-borne detectors, or from indirect measurements by air shower detector arrays on the Earth's surface. Imaging Atmospheric Cherenkov Telescopes (IACTs), used primarily for $\gamma$-ray astronomy, can also be used for cosmic-ray physics. They are able to measure Cherenkov light emitted both by heavy nuclei and by secondary particles produced in air showers, and are thus sensitive to the charge and energy of cosmic ray particles with energies of tens to hundreds of \TeV. A template-based method, which can be used to reconstruct the charge and energy of primary particles simultaneously from images taken by IACTs, will be introduced. Heavy nuclei, such as iron, can be separated from lighter cosmic rays with this method, and thus the abundance and spectrum of these nuclei can be measured in the range of tens to hundreds of \TeV.
}

\FullConference{The 34th International Cosmic Ray Conference,\\
		30 July -- 6 August, 2015\\
		The Hague, The Netherlands}

\begin{document}

\section{Introduction}
Cosmic rays (CRs) are charged particles of extraterrestrial origin impinging on the Earth's atmosphere. They are mostly made up of protons, with a small fraction of other fully ionized nuclei, electrons, and antiparticles. The CR energy spectrum is smooth and almost featureless from a few \GeV to several \exa\eV{}. It can be described by a power law with two breaks: the \textit{knee} (steepening at a few \peta\eV{}) and the \textit{ankle} (flattening at a few \exa\eV{}).

\begin{figure}[t]
\begin{center}
\includegraphics[width=10cm, trim=19cm 0.7cm 10cm 1.5cm, clip]{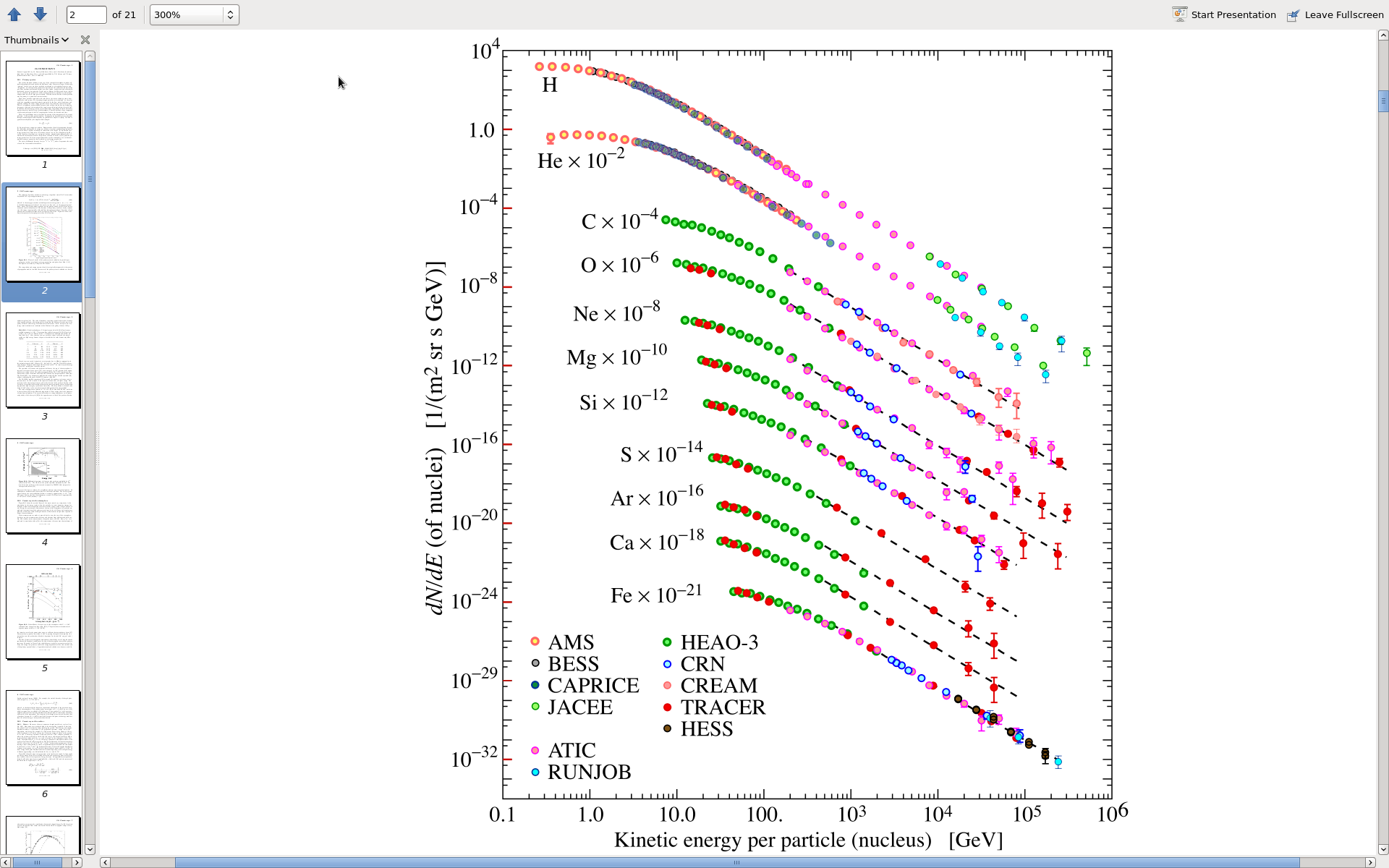}
\caption{\label{pdgspectrum}Energy spectra of various elements in cosmic rays, measured by different experiments over a large energy range, from \cite{PDG-2012}.}
\end{center}
\end{figure}

CRs are deflected by (extra-)galactic magnetic fields, making it impossible to identify their sources (acceleration sites) directly for the energies considered here. However, we can identify acceleration sites by studying neutral byproducts of CR acceleration, such as $\gamma$-rays and neutrinos. Another way to study the sources of cosmic rays is to precisely measure their spectrum and composition, and compare these to the predictions from models of different types of acceleration sites, see e.g. \citet{knee}. Examples of such candidate acceleration sites are supernova remnants for galactic cosmic rays and gamma-ray bursts or active galactic nuclei for extra-galactic cosmic rays. Some of the results of these precision measurements are shown in Figure \ref{pdgspectrum}. 

\subsection{Experimental Methods in Cosmic Ray Physics}
In the \mega\eV{} to \TeV{} range, cosmic rays can be measured directly by balloon- or satellite-borne detectors placed at the top of the atmosphere, see for example \citet{ahn2009}. The energy, mass and charge of the cosmic ray can be determined directly, from the energy deposited in the detector and the shape of the particle's track.

Upon interaction with the Earth's atmosphere, cosmic rays with energies above a few hundred \TeV{} will produce a cascade or \textit{shower} of secondary particles (nucleons, electrons, positrons, pions, muons, photons, neutrinos). If the primary particle's energy is high enough, part of this \textit{extensive air shower} (EAS) may be detected by arrays of particle detectors on the ground. Energy and direction of the primary particle can be inferred from the density and arrival times of the secondaries. 

For intermediate energy primaries, no, or only a few, secondary particles will reach the ground. However, the charged component of the shower will emit Cherenkov radiation in the visible/UV range, high up in the atmosphere. This light can be detected by Imaging Atmospheric Cherenkov Telescopes (IACTs), mostly used for $\gamma$-ray astronomy. These instruments use very sensitive cameras to image air showers and reconstruct energy, direction, and type of the primary particle from the brightness, shape, and orientation of the image. Using the direct Cherenkov technique \citet{Kieda} described below, they can measure the charge of the primary particles, especially for heavy nuclei. This enables the use of large existing data sets for cosmic ray physics (typically up to 1000 hours of quality-selected data per year of operation, with a few hundred cosmic ray events per second, which are considered as background for $\gamma$-ray astronomy). The systematic uncertainties are dominated by the atmosphere and thus largely complementary to the direct detection and EAS experiments.

\subsection{Direct Cherenkov Technique}
In addition to the Cherenkov light radiated by the charged component of the shower, charged primaries with high velocities will also radiate direct Cherenkov (DC) light before starting a shower. The DC light is emitted at high altitudes under very small angles. For today's generation of IACTs, with typical pixel sizes on the order of \unit{0.15}{\degree} \citet{veritas}, this light is generally concentrated within one (or two neighboring) pixel(s). The intensity is proportional to $Z^2$, the squared charge of the primary nucleus. In combination with an energy estimate from the intensity of the shower image, the DC light can be used to obtain separate spectra for light and heavy nuclei if the DC contribution can be identified as such and distinguished from the light emitted by the secondary particles. Light elements such as helium or oxygen are not very well suited for the direct Cherenkov technique as the DC light emitted by these elements is too dim compared to the contribution from the air shower. For example, iron emits ten times as much DC light as oxygen ($Z=8$). In the following, iron $(Z=26)$ will be used as a representative for heavy elements as it is the most abundant element in cosmic rays with $Z>20$. 

The DC technique is sensitive to iron nuclei in cosmic rays with energies of about \unit{10}{\TeV} to several \unit{100}{TeV}. For lower energies, the particle will interact and start a shower before emitting Cherenkov light. At higher energies, the light emitted by the shower is brighter than the DC contribution. For more details, see \cite{rolfpaper, Kieda, wissel}.

\subsection{The VERITAS Experiment}
VERITAS (Very Energetic Radiation Imaging Telescope Array System) \cite{veritas,naheeperformance} is an array of four IACTs located at the Fred Lawrence Whipple Observatory (FLWO) in southern Arizona (31 40N, 110 57W,  1.3km a.s.l.). It has been operational since 2007. Each telescope has a total mirror area of $\sim$\unit{100}{\square\meter}, and a camera consisting of 499 \textit{pixels} (photo multiplier tubes), each with a field of view of \unit{0.15}{\degree} diameter. It is sensitive to showers induced by $\gamma$-rays from \unit{85}{\GeV} to $>$ \unit{30}{\TeV}. 

The standard data analysis chain is described in \citet{veritasanalysis, vegas}. After charge integration and image cleaning, a moments analysis is performed and the \textit{Hillas parameters} \citet{hillas} are calculated. From the orientation and shape of the images, the direction of the primary particles and the position of the shower core on the ground can be determined, and $\gamma$-ray induced showers separated from cosmic-ray induced ones. Look-up tables are used to determine the energy given the intensity of the shower and the distance to the telescope. These results may be used directly to determine spectra, fluxes, etc, or used as input to more advanced reconstruction techniques such as the template likelihood fitting method described below.

\section{Template Likelihood analysis method}

This reconstruction method requires a model that predicts the average photon intensity in the camera, depending on some properties of the primary particle such as energy, arrival direction, etc. The probability distribution of the signal (charge) per camera pixel is then determined by those parameters as well as the detector response and its uncertainty. Given the images taken of a particular shower, this probability distribution can be converted into a likelihood function of the primary parameters. Maximizing this likelihood function produces a set of parameters which are good estimates for the ``true'' properties of the primary particle. In addition, the goodness-of-fit for those parameters can be used to separate background events (that are not described by the model) from signal events (that are well-described by the model). 

A similar likelihood fitting method has been used for the reconstruction of $\gamma$-ray induced showers by other experiments, see for example \cite{cattemplate}, as well as by VERITAS \citet{frogs}, to improve the performance of the experiment.

In this paper, we describe the performance of a variant of the likelihood fitting method  used to reconstruct the energy of cosmic ray irons from their associated air showers, and to separate iron-induced air showers from those induced by lighter cosmic rays. The model parameters are the primary particle's energy $E$, arrival direction in camera-centered coordinates ($X_s$, $Y_s$), height of first interaction $h$ and position of the shower core on the ground ($X_p$, $Y_p$). The light intensity in the camera is modeled as a set of \textit{templates}, using Monte Carlo simulations of iron-induced showers in a fixed grid in $E$, $h$, and the distance between the telescope and the shower core. Several simulated showers are averaged to obtain a stable average of the photon intensity in the camera. Interpolation is used to predict the light intensity for arbitrary parameter values. Rotation and translation of the template image account for different arrival directions and core positions on the ground.

\begin{figure}[t]
\begin{center}
\begin{overpic}[width=25pc]{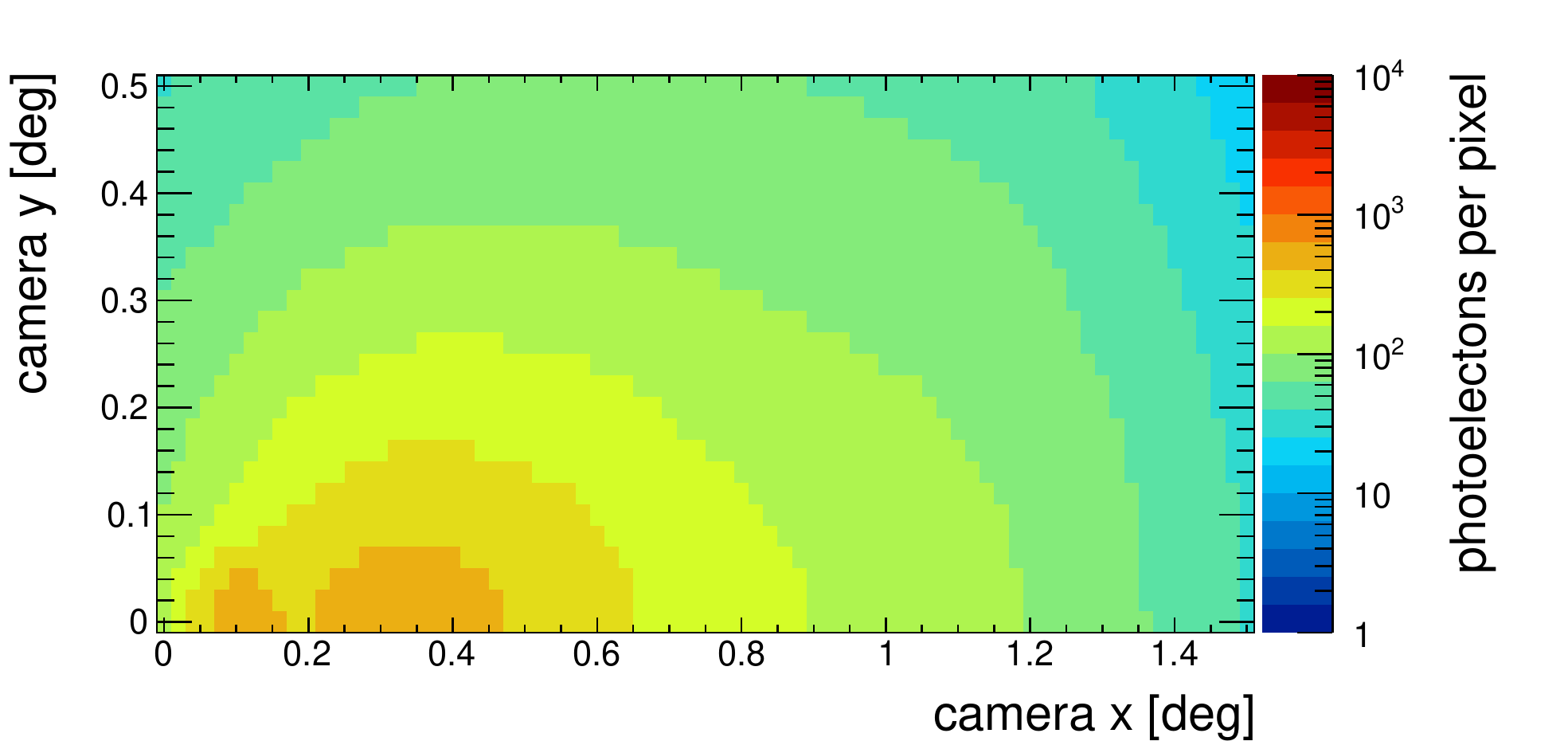}
\thicklines
\put( -10, 10) {\fbox{\textsf{\textbf{DC light}}}}
\put( 30, 15) {\fbox{\textsf{\textbf{shower light}}}}
\put( 7,9) {\vector(1,0){8}}
\put( 52, 44) {\textsf{\textbf{\footnotesize{VERITAS --- ICRC 2015}}}}
\end{overpic}
\caption{Predicted average light distribution in the camera (\textit{template}) for iron showers with $E=\unit{30}{\TeV}$, with a distance of \unit{50}{\metre} between the detector and shower core, $h=\unit{33}{\kilo\metre}$. The origin is defined by the primary particle's arrival direction. The x-axis is defined by the long image axis. Note the contribution from direct Cherenkov light at around \unit{0.1}{\degree} from the primary's direction, clearly distinguishable from the broader contribution from the Cherenkov light emitted by the shower particles. 
}
\label{template}
\end{center} 
\end{figure}

CORSIKA \cite{corsika} was used for air-shower simulation including the emission of Cherenkov light, and the grisudet package\footnote{\url{http://www.physics.utah.edu/gammaray/GrISU/}} was used for ray-tracing in the telescopes. Figure \ref{template} shows an example of the predicted average light intensity in the camera for an iron shower. The contributions from both DC light and the air shower can be clearly seen. A more detailed description of the reconstruction process, can be found in \citet{frogs}. The likelihood function adapted for the reconstruction of iron showers is described in \citet{ECRSiron}.

\begin{figure}[f]
\begin{center}
\begin{minipage}[t]{0.5\textwidth}
\begin{overpic}[width=0.85\textwidth]{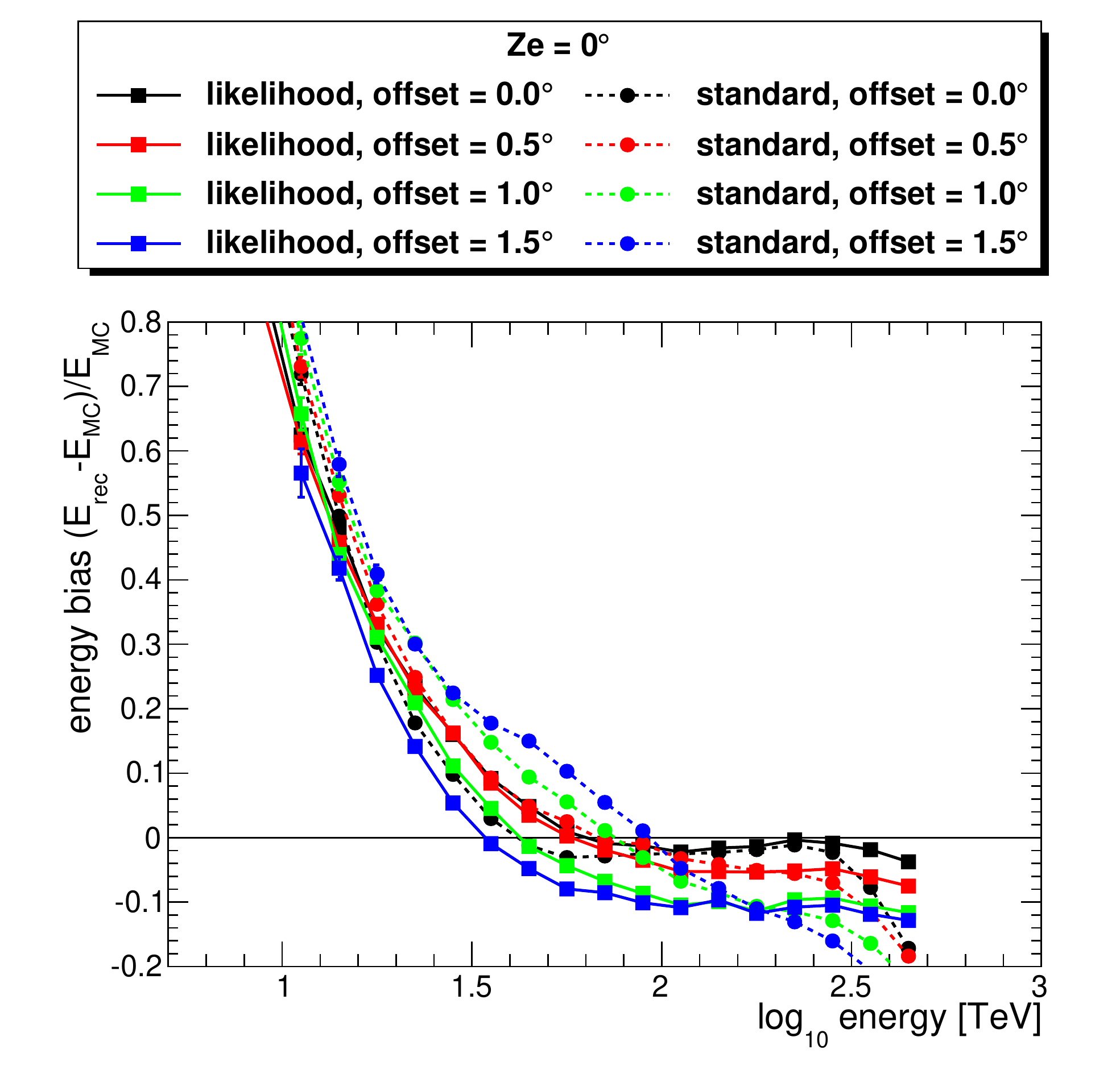}
\put( 37, 60) {\textsf{\textbf{\footnotesize{VERITAS --- ICRC 2015}}}}
\end{overpic}
\caption{Energy bias for \unit{0}{\degree} zenith angle.}\label{ebias2}
\vspace{0.45cm}
\begin{overpic}[width=0.85\textwidth]{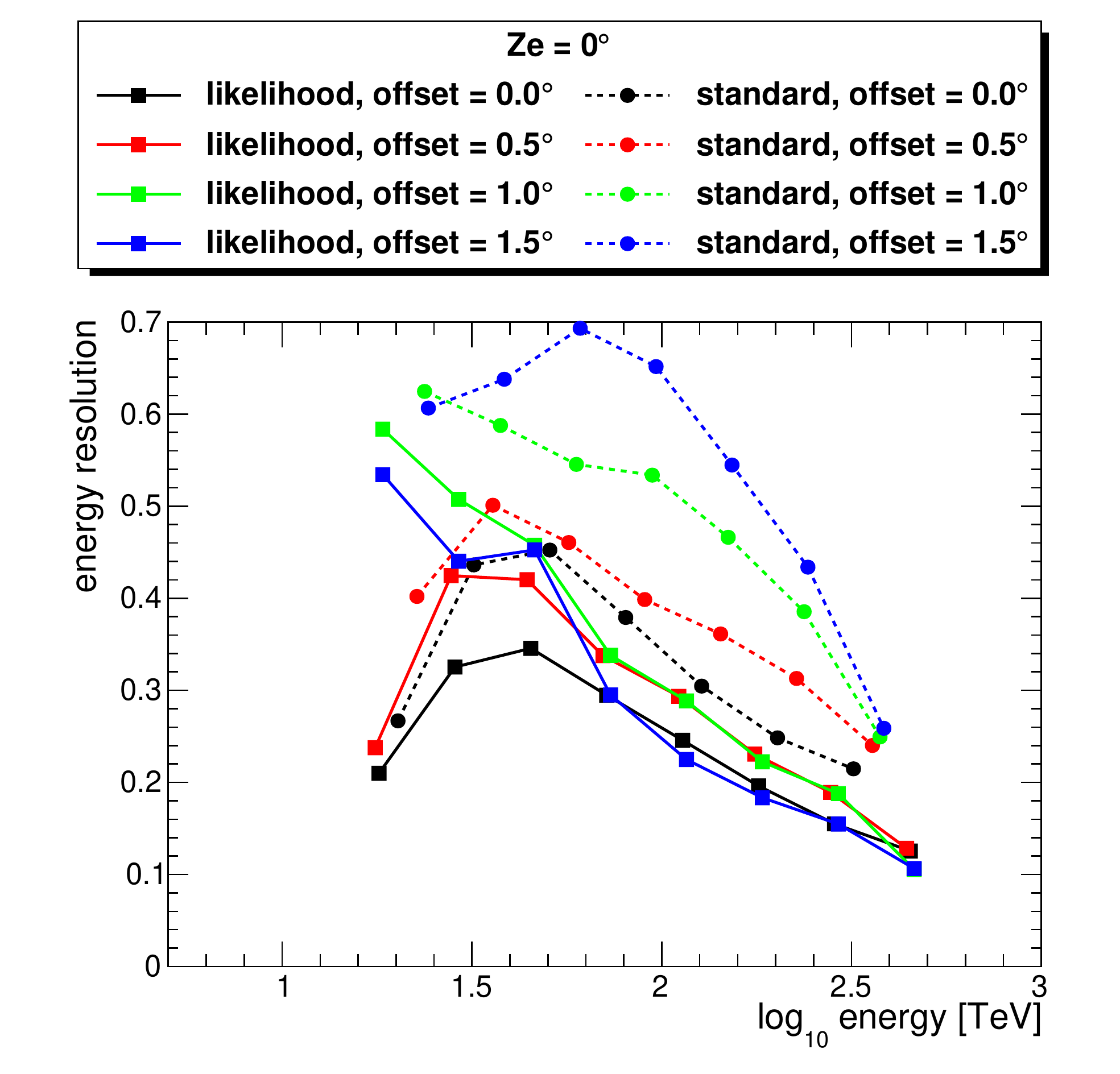}
\put( 37, 12) {\textsf{\textbf{\footnotesize{VERITAS --- ICRC 2015}}}}
\end{overpic}
\caption{Energy resolution for \unit{0}{\degree} zenith angle.}\label{eres2}
\vspace{0.45cm}
\begin{overpic}[width=0.85\textwidth]{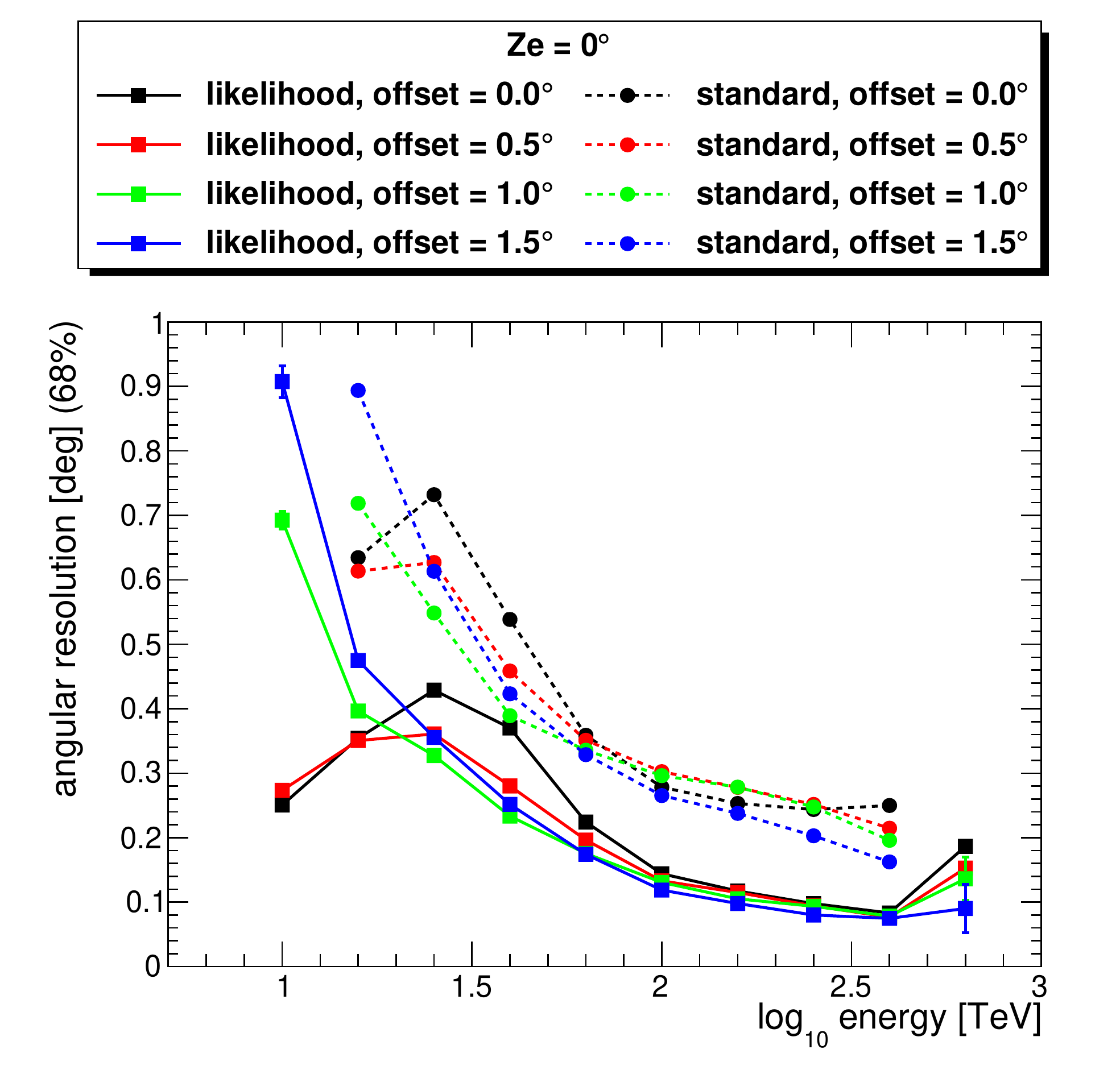}
\put( 37, 60) {\textsf{\textbf{\footnotesize{VERITAS --- ICRC 2015}}}}
\end{overpic}
\caption{Angular resolution for \unit{0}{\degree} zenith angle.}\label{corres2}
\end{minipage}\hspace{-0pc}%
\begin{minipage}[t]{0.5\textwidth}
\begin{overpic}[width=0.85\textwidth]{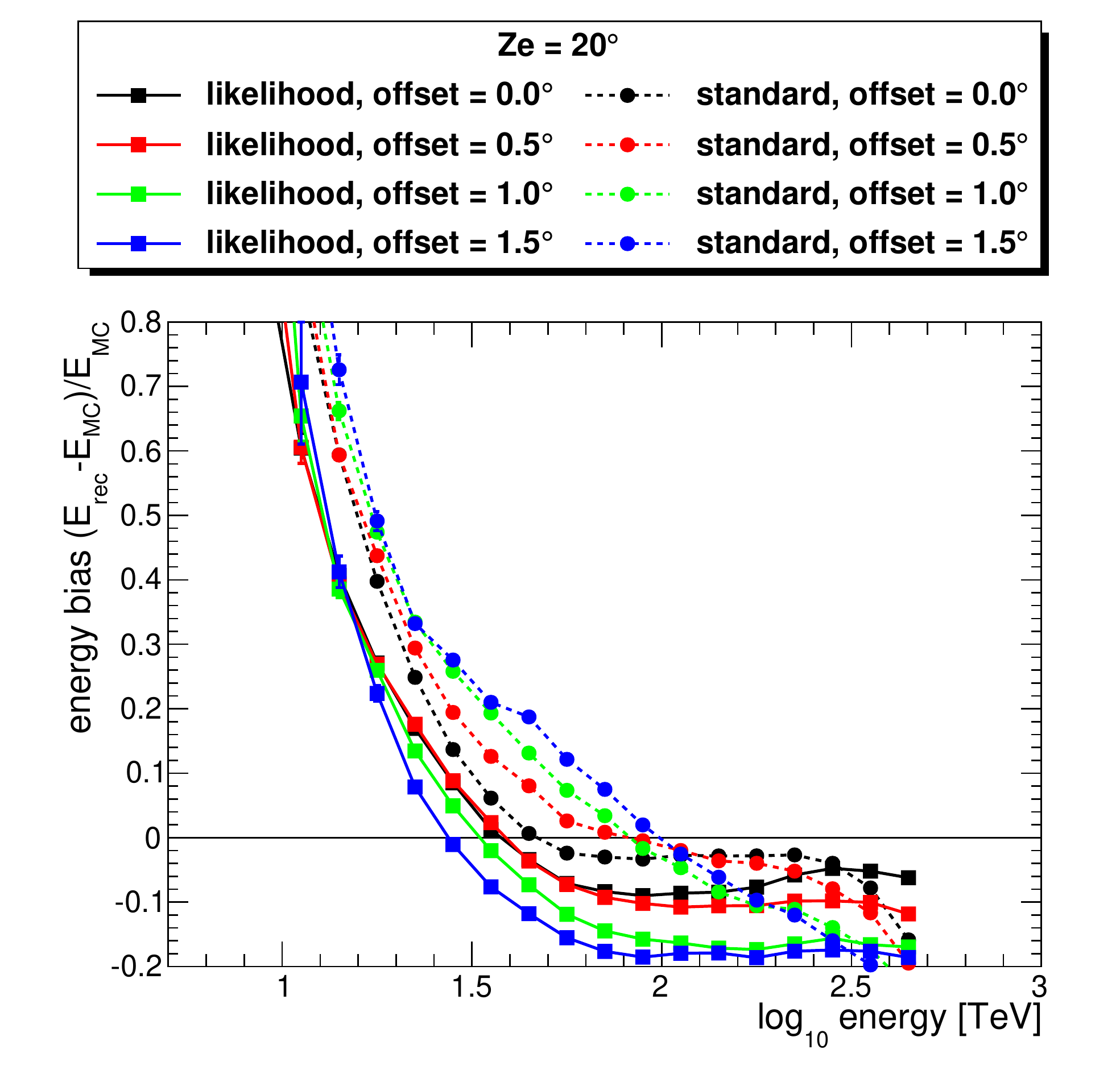}
\put( 37, 60) {\textsf{\textbf{\footnotesize{VERITAS --- ICRC 2015}}}}
\end{overpic}
\caption{Energy bias for \unit{20}{\degree} zenith angle.}\label{eres5}
\vspace{0.45cm}
\begin{overpic}[width=0.85\textwidth]{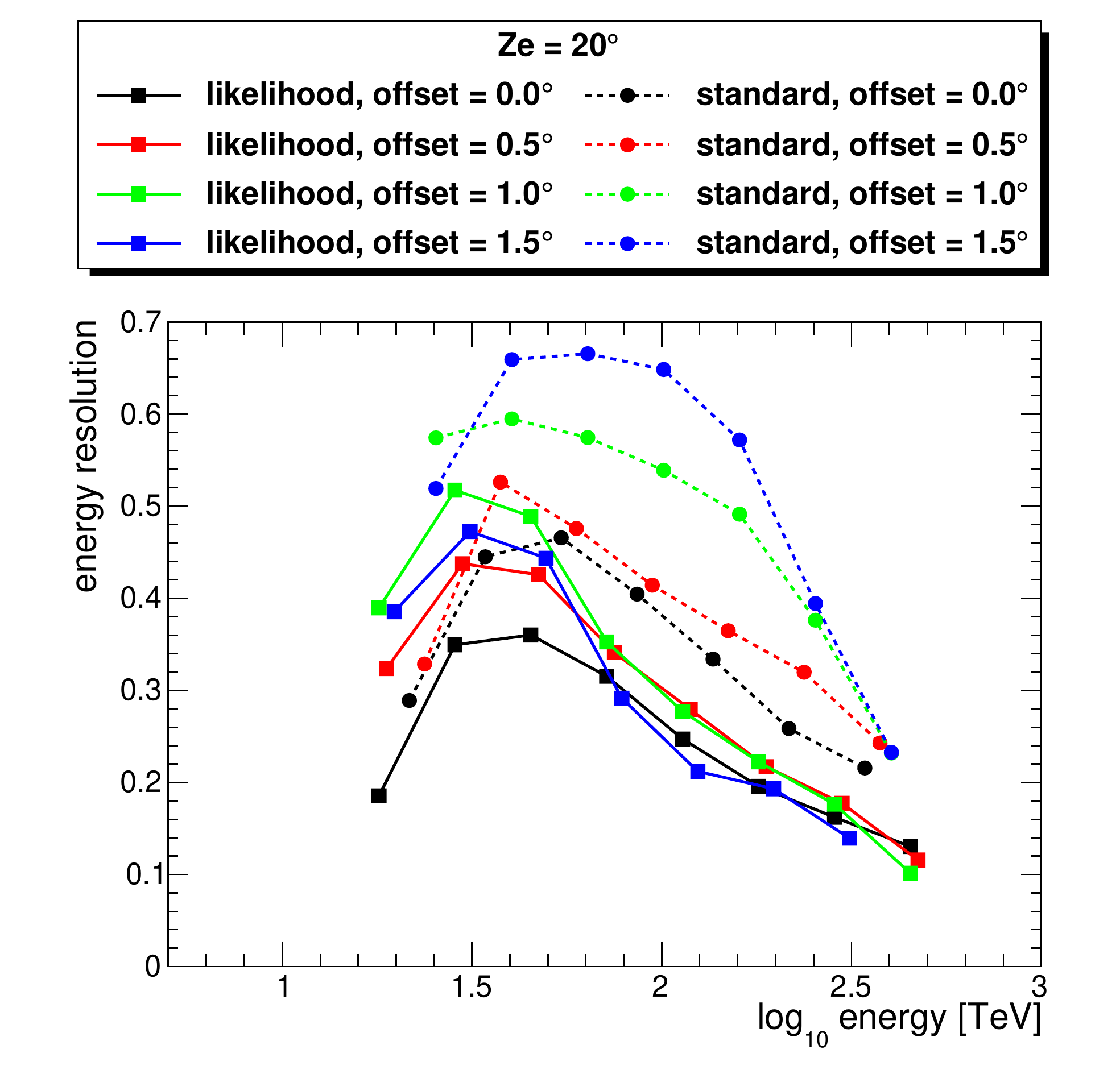}
\put( 37, 12) {\textsf{\textbf{\footnotesize{VERITAS --- ICRC 2015}}}}
\end{overpic}
\caption{Energy resolution for \unit{20}{\degree} zenith angle.}\label{ebias5}
\vspace{0.45cm}
\begin{overpic}[width=0.85\textwidth]{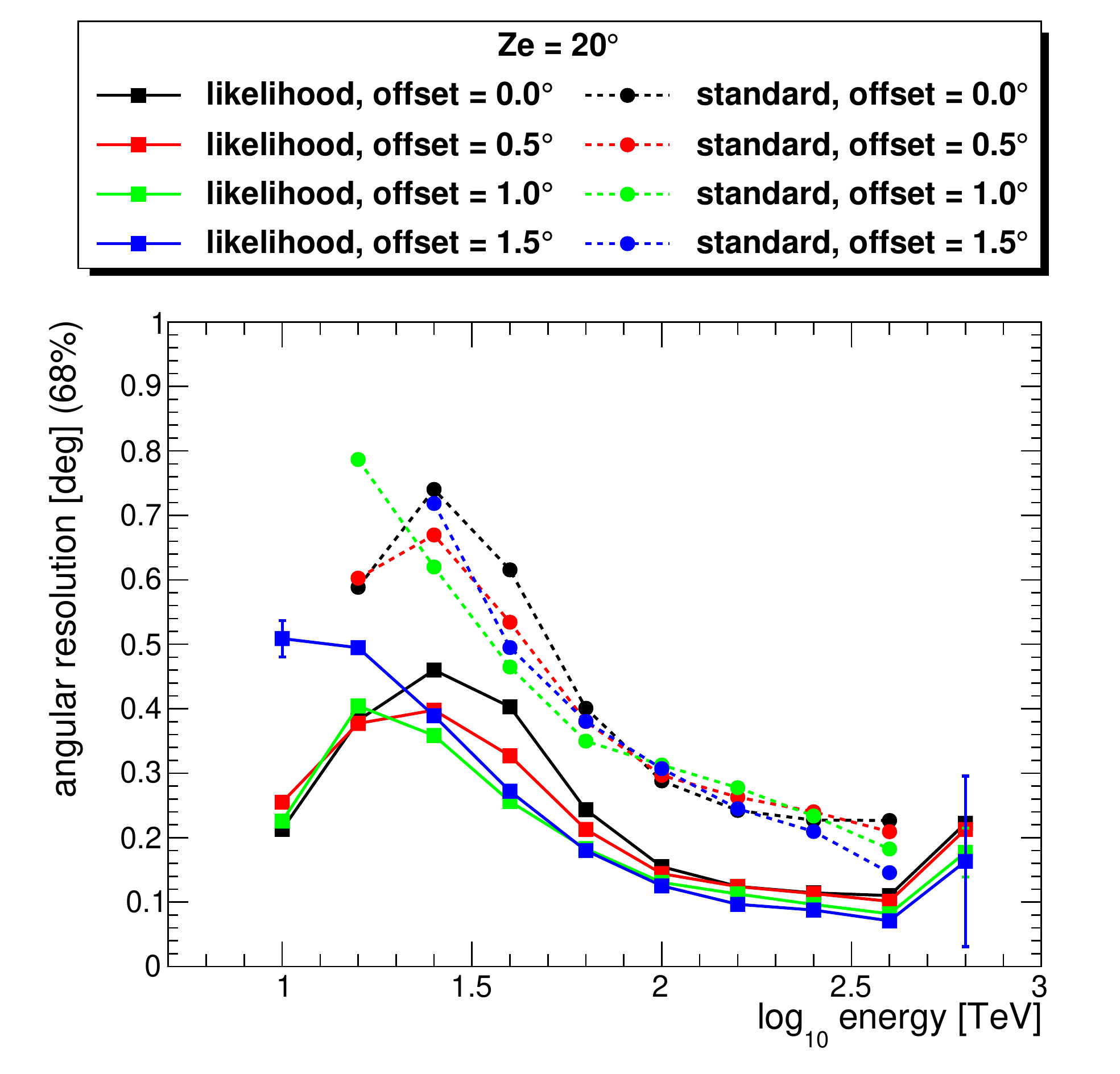}
\put( 37, 60) {\textsf{\textbf{\footnotesize{VERITAS --- ICRC 2015}}}}
\end{overpic}
\caption{Angular resolution for \unit{20}{\degree} zenith angle.}\label{corres5}
\end{minipage}
\end{center} 
\end{figure}

\section{Method Performance}

To test the method, a set of templates was produced using simulated iron showers from zenith in the energy range from 10 -- \unit{100}{\TeV}. These templates were tested against a set of simulated iron showers with random energies, core positions and arrival directions in the camera, which was passed through a simulation of the full VERITAS detector, again using the grisudet package. These showers were analyzed with the template fitting method described earlier. Starting values for the template fit were obtained using the standard geometrical reconstruction, with special lookup tables for iron-induced showers. 

The following three quantities are plotted in Figures \ref{ebias2} -- \ref{corres5} and compared: The energy bias, defined as the relative error in the energy $\frac{E_\mathrm{rec}}{E_\mathrm{MC}}-1$, the energy resolution ($68\%$ containment interval around the median reconstructed energy), and the angular resolution ($68\%$ containment radius of the reconstructed direction around the true direction). Energy bias and resolution are important sources of systematic uncertainties on the energy spectrum. The angular resolution is plotted here to show that the direction is reconstructed well enough to identify the DC light contribution. All quantities are plotted both for the template likelihood fitting reconstruction (``likelihood'', solid lines) and the standard reconstruction (``standard'', dashed lines). The template method always performs better than the geometric reconstruction. Quality cuts are applied, but no cut on the goodness-of-fit after the likelihood minimization. Note that the performance of the standard reconstruction is worse for iron-induced showers than for $\gamma$-ray induced showers (as described in \citet{naheeperformance}). This is due to the fact that hadronic showers tend to be much less smooth, with larger shower-to-shower variations than $\gamma$-ray induced showers.

Figure \ref{ebias2} shows the energy bias depending on the arrival direction (offset from the camera center), with the telescopes pointing at zenith. Above \unit{50}{\TeV}, it the energy bias is very flat. There is a significant dependence on the offset from the camera center: For large offsets, the bias is as large as $10\%$. This happens because the templates were produced for zero offset. The VERITAS optical point-spread function degrades with the offset, so there is less light and hence the energy is underestimated. Since the energy bias is relatively flat, this is can be corrected for in the future. 

Figures \ref{eres2} and \ref{corres2} show the energy resolution and angular resolution for different offsets. The resolution is significantly better than for the geometric reconstruction, and the dependence of the energy resolution on the offset it greatly reduced. However, below about \unit{80}{\TeV}, the angular resolution is still larger than the pixel diameter, which implies that the DC light, which should be contained within one pixel, is not always identified properly.

Figures \ref{eres5}, \ref{ebias5}, and \ref{corres5} show the corresponding plots for the telescopes pointing \unit{20}{\degree} away from zenith, averaged over all azimuth directions. The energy bias is slightly worse, due to the fact that the light had to travel longer through the atmosphere, so more light was absorbed. Energy and angular resolution are not degraded. The dependence on the azimuth was investigated as well (not shown). There is no significant dependence of the energy bias, energy resolution or angular resolution on the azimuth.

\section{Conclusions and Outlook}
To fully understand the sources and acceleration mechanisms of cosmic rays, it is important to measure their composition and spectrum over a large energy range. The range of tens to hundreds of TeV is not well-covered by existing direct detection experiments. However, IACTs are sensitive to cosmic rays in that energy range. By detecting direct Cherenkov light emitted high in the atmosphere, they can separate light and heavy cosmic rays and produce separate spectra. 

A method for reconstructing the energy of iron-induced showers, imaged by the VERITAS experiment, has been presented. This likelihood fitting technique relies on fitting template images to recorded data, and can be used to separate iron-induced showers from background events as well. However, further work is needed to improve the performance of the fit and to optimize the separation between showers induced by iron and those induced by light nuclei.

In the future, this method will be applied to archival data sets taken with the VERITAS instrument. The goal is to obtain an energy spectrum of iron nuclei in cosmic rays for energies in the range of about 30 to \unit{300}{\TeV}. Due to the improved reconstruction method, this is expected to reduce the uncertainty on the spectral measurements compared to current results.

\acknowledgments
This research is supported by grants from the U.S. Department of Energy Office of Science, the U.S. National Science Foundation and the Smithsonian Institution, and by NSERC in Canada. We acknowledge the excellent work of the technical support staff at the Fred Lawrence Whipple Observatory and at the collaborating institutions in the construction and operation of the instrument. The VERITAS Collaboration is grateful to Trevor Weekes for his seminal contributions and leadership in the field of VHE gamma-ray astrophysics, which made this study possible.

H.F. acknowledges support through the Helmholtz Alliance for Astroparticle Physics.

\clearpage

\bibliography{bib}

\end{document}